\documentclass[dvips]{article} 
\usepackage{spie}   
\usepackage{psfig}

\title{Exploration of Parameter Spaces in a Virtual Observatory} 

\author{
S.G. Djorgovski\supit{a},
A. Mahabal\supit{a},
R. Brunner\supit{a},
R. Williams\supit{a},
\skiplinehalf
R. Granat\supit{b},
D. Curkendall\supit{b},
J. Jacob\supit{b},
P. Stolorz\supit{b}
\skiplinehalf
\supit{a} Palomar Observatory, Caltech, Pasadena, CA, 91125, USA
\skiplinehalf
\supit{b} Jet Propulsion Laboratory, Pasadena, CA, 91109, USA
}
\authorinfo{
{\bf To appear in Proc. SPIE v. 4477 (2001)}
}

\begin{document} 
  \maketitle 


\begin{abstract}
Like every other field of intellectual endeavor, astronomy is being
revolutionised by the advances in information technology. There is an ongoing
exponential growth in the volume, quality, and complexity of astronomical data
sets, mainly through large digital sky surveys and archives.  The Virtual
Observatory (VO) concept represents a scientific and technological framework
needed to cope with this data flood.  Systematic exploration of the observable
parameter spaces, covered by large digital sky surveys spanning a range of
wavelengths, will be one of the primary modes of research with a VO.  This is
where the truly new discoveries will be made, and new insights be gained about
the already known astronomical objects and phenomena.  We review some of the
methodological challenges posed by the analysis of large and complex data sets
expected in the VO-based research. The challenges are driven both by the size
and the complexity of the data sets (billions of data vectors in parameter
spaces of tens or hundreds of dimensions), by the heterogeneity of the data and
measurement errors, including differences in basic survey parameters for the
federated data sets (e.g., in the positional accuracy and resolution,
wavelength coverage, time baseline, etc.), various selection effects, as well
as the intrinsic clustering properties (functional form, topology) of the data
distributions in the parameter spaces of observed attributes.  Answering these
challenges will require substantial collaborative efforts and partnerships
between astronomers, computer scientists, and statisticians. 

\end{abstract}



\section{Introduction: The Virtual Observatory Concept}
\label{sect:intro}  

Observational astronomy is undergoing a paradigm shift.  This revolutionary
change is driven by the enormous technological advances in telescopes and
detectors (e.g., large digital arrays), the exponential increase in computing
capabilities, and the fundamental changes in the observing strategies used to
gather the data.  In the past, the usual mode of observational astronomy was
that of a single astronomer or small group performing observations of a small 
number of objects (from single objects and up to some hundreds of objects).
This is now changing: large digital sky surveys over a range of wavelengths,
from radio to x-rays, from space and ground are becoming the dominant source 
of observational data.  Data-mining of the resulting digital sky archives is
becoming a major venue of the observational astronomy.  The optimal use of
the large ground-based telescopes and space observatories is now as a follow-up
of sources selected from large sky surveys.  This trend is bound to continue,
as the data volumes and data complexity increase.  The very nature of the
observational astronomy is thus changing rapidly.

The existing surveys already contain many Terabytes of data, from which
catalogs of many millions, or even billions of objects are extracted.  For each
object, some tens or even hundred parameters are measured, most (but not all)
with quantifiable errors.  Some of these data sets are described by
Brunner {\it et al.} (2001b).
Forthcoming projects and sky surveys are expected to deliver data volumes
measured in Petabytes. 
A major new area for exploration will be in the time domain
(cf. Paczy\'nski 2000), with a number of ongoing or forthcoming surveys aiming
to map large portions of the sky in a repeated fashion, down to very faint flux
levels.  These synoptic surveys will be generating Petabytes of data, and they
will open a whole new field of searches for variable astronomical objects. 

Every astronomical observation and every survey covers a portion of the
observable parameter space, whose axes include the area coverage, wavelength
coverage, limiting flux, etc., and with a limited resolution in angular scales,
wavelength, temporal baseline, etc.  Each one represents a partial projection
of the observable universe, limited by the observational or survey parameters
(e.g., pixel sampling, beam size, filters, etc.).  Federating multiple
surveys sampling different portions of the observable parameter space thus
provides a much more complex, but also a more complete view of the physical
universe. 

This great richness of information is hard to translate into a derived
knowledge and physical understanding.  Questions abound: 
How do we effectively explore datasets comprising hundreds of millions or
billions of objects each with hundreds of attributes? 
How do we objectively classify the detected sources to isolate subpopulations
of astrophysical interest, and quantify their properties? 
How do we identify correlations and anomalies within the data sets?  
How do we use the data to constrain astrophysical interpretation, which often
involve highly non-linear parametric functions derived from fields such as
physical cosmology, stellar structure, or atomic physics? 
How do we match these complex data sets with equally complex numerical
simulations, and how do we evaluate the performance of such models?
And equally important, what are the limits of our current knowledge imposed
by the existing data?
 
The key task is now to enable an efficient and complete scientific exploitation
of these enormous data sets.  Similar situations exist in many other fields of
science and applied technology today.  This poses many technical and conceptual
challenges, and it may lead to a whole new methodology of doing science in the
information-rich era. 

In order to cope with this data flood, the astronomical community started a
grassroots initiative, the National (and ultimately Global) Virtual Observatory.
A Virtual Observatory would federate numerous large digital sky archives,
provide the information infrastructure and standards for ingestion of new data
and surveys, and develop the computational and analysis tools with which to
explore these vast data volumes.  Responding to this urgent need, the National
Academy of Science Astronomy and Astrophysics Survey Committee, in its new
decadal survey, {\em Astronomy and Astrophysics in the New Millennium} 
(McKee, Taylor, {\it et al.} 2001)
recommends, as a first priority, the establishment of a National Virtual
Observatory (NVO).
Recognising the ultimately international/global nature of this concept, we will
hereafter refer to it simply as a Virtual Observatory (VO).

The VO would provide new opportunities for scientific discovery that were
unimaginable just a few years ago.  Entirely new and unexpected scientific
results of major significance will emerge from the combined use of the
resulting datasets, science that would not be possible from such sets used
singly.  In the words of a White Paper (2001),
the VO will serve as {\em an engine of discovery for astronomy.}

Implementation of the VO involves significant technical challenges on many
fronts, including database architecture, archive federation, standardisation
of metadata, data formats, and exchange protocols, and above all,
{\em data analysis and understanding}.  This will include data mining,
visualisation, novel statistical techniques, etc. 

A number of reviews and contributed papers relevant for the subject can be
found in the volumes edited by Brunner {\it et al.} (2001) and Banday {\it et
al.} (2001), and in these Proceedings. 
Here we review some of the novel scientific directions and technical challenges
posed by a general research function of the VO, the exploration of large data
parameter spaces derived from massive digital sky surveys.

\section{Exploration of Parameter Spaces: A Systematic Approach to Discovery} 
\label{parspac}  

Every astronomical data set, from individual measurements, images, or spectra,
to large digital sky surveys, covers some finite portion of the observable
parameter space, and is thereby limited in its descriptive power.

Some axes of the observable parameter space are obvious and well understood:
the flux limit (depth), the solid angle coverage, and the range of wavelengths
covered (or spectra).  Others include the limiting surface brightness (over a
range of angular scales), angular resolution, wavelength resolution,
polarization, and especially variability over a range of time scales (or more
generally, temporal power spectra); all of them applying at any wavelength, and
again as a function of the limiting flux.  
In some cases (e.g., the Solar system, Galactic structure) apparent and proper
motions of objects are detectable, adding additional information axes.  For
well-resolved objects (e.g., galaxies), there should be some way to quantify
the image morphology as one or more parameters.  And then, then there are the
non-electromagnetic information channels, e.g., neutrinos, gravity waves,
cosmic rays \ldots The observable parameter space is enormous!

Every astronomical observation or data set, large digital sky surveys included,
samples only a small portion of this grand observable parameter space, usually
covering only some of the axes and only with a limited dynamical range along
each axis.  Every survey is also subject to its own selection and measurement
limits, e.g., limiting fluxes, surface brightness, angular resolution,
spectroscopic resolution, sampling and baseline for variability if multiple
epoch observations are obtained, etc. 
Surveys thus represent hypervolumes in the observable parameter space,
delimited by the survey parameters.
Individual sources represent data points (or vectors) in this multidimensional
space, and are detected if they fall within one or more of the surveyed
volumes.

We can thus, in principle, measure a huge amount of information arriving from
the universe, and so far we have sampled well only a relatively limited set of
sub-volumes of this large, observable parameter space.  We have a much better 
coverage along some axes than others: for example, we have fairly good sky
surveys in the visible, NIR, and radio; more limited all-sky surveys in the
x-ray and FIR regimes; etc.  Likewise, some gaps in the coverage are glaring:
for example, it would be great to have an all-sky survey at the FIR and sub-mm 
wavelengths, reaching to the flux levels we are accustomed to in the visible or
radio surveys, and with an arcsecond-level angular resolution; this is
currently technically difficult and prohibitively expensive, but it is
possible.  The whole time domain is another great potential growth area.  Some
limits are simply technological or practical (e.g., the cost issues); but some
are physical, e.g., the quantum noise limits, or the opacity of the Galactic
ISM.

Historically, the concept of the systematic exploration of the universe through
a systematic study of the observable parameter space was pioneered by Fritz
Zwicky, starting in 1930's (see, e.g., Zwicky 1957).  While his methodology and
approach did not find many followers, the core of the important ideas was
clearly there.  Zwicky was limited by the technology available to him at the
time; probably he would have been a major developer and user of a VO today! 
Another interesting approach was taken by Harwit (1975; see also Harwit \&
Hildebrand 1986), who examined the limits and selection effects operating on a
number of axes of the observable parameter space, and tried to estimate the
number of fundamental new (``class A'') astrophysical phenomena remaining to be
discovered.  While one could argue with the statistics, philosophy, or details
of this analysis, it poses some interesting questions and offers a very general
view of our quest to understand the physical universe. 

So, it is not just the space we want to study; it is the observable parameter 
space, as a means of encoding systematically our empirical knowledge about the
universe.  Much of the total observable parameter space which is in principle
(i.e., technologically) available to us is still very poorly sampled: astronomy
is far from being a finished science!

The unexplored regions of the parameter space are our {\sl Terra Incognita},
which we should explore systematically, and where we have our best chance to
uncover some previously unknown types of objects or astrophysical phenomena ---
as well as reach a better understanding of the already known ones.  This is an
ambitious, long-term program, but even with a relatively limited coverage of
the observable parameter space we already have in hand it is possible to make
some significant advances. 

A complete parameter space representation of our measured coverage of the
universe is naturally enabled by a VO.  Such a broad picture can then serve not
only as a {\it framework for scientific investigations,}
but also as a {\it tool of scientific planning and strategy}:  
Where do we know the least about the universe?  
Where should we position the next sky survey or the space mission? 
How can we most effectively bridge the gaps in our complex, multi-dimensional
view of the universe?  
Having a good, top-level picture of the domains of our knowledge (or the lack
thereof) seems necessary in order to make truly informed decisions about the
scientific exploration strategy. 

We now address some specific issues and approaches to this ambitious scientific
program.

\section{Constructing a Panchromatic Universe: Survey Federation} 
\label{surfed}  

The first issue we are facing in assembling of VO data sets is that of
survey federation (sometimes also called the data fusion).  This would 
generally produce an immediate added-value benefits within a VO, since a
broader baseline in wavelength for combination of surveys done at different
wavelengths (a panchromatic universe) or in time, for a combination of surveys
done at the same wavelengths but at different epochs (a synoptic universe),
inevitably contains additional information beyond what is available in any
of the surveys taken separately.  Reliability and accuracy of source matching
in this process are thus critical for the scientific uses of federated datasets.

\begin{figure}
\begin{center}
\begin{tabular}{c}
\psfig{figure=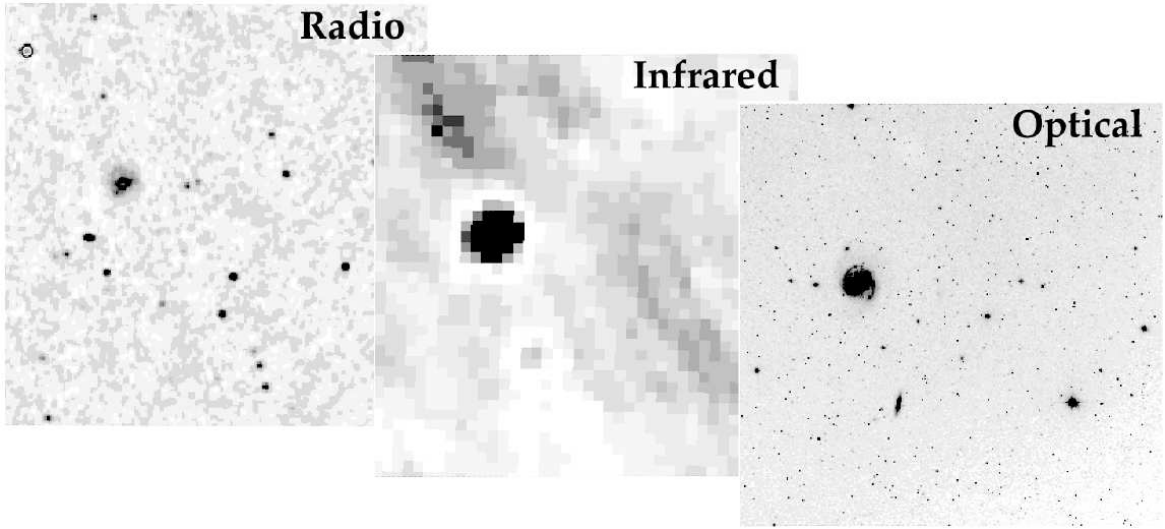,height=7cm}
\end{tabular}
\end{center}
\caption[figure 1] 
{ \label{fig:one}
A schematic illustration of the problem of a panchromatic survey federation,
where individual surveys may differ widely in terms of the angular resolution,
source number density, foregrounds, and S/N ratios.  If multiple positional
coincidences for a given source ID are possible, additional information must
be used to discern the most probable match.
}
\end{figure} 

The sky is surveyed over a range of wavelengths, and to the flux and surface
brightness limits, each survey has its own limiting angular resolution: in the
optical and NIR, this is usually given by the seeing (the atmosphere), and at
most other wavelengths by the telescope beam size.  Survey pixel sampling rates
are adjusted accordingly.  The beam size and the S/N ratio determine the
accuracy of the source positions; their absolute astrometry (conversion to a
standard cellestial sphere coordinate system) introduces additional systematic
uncertainties. 

Federation of surveys is accomplished by cross-identification of sources
detected in each one (however, many sources may be detected in only one
survey or a bandpass).  If the positional accuracies and/or the beam sizes
for the surveys are comparable, simple positional matching is usually
sufficient; but if the effective angular resolutions and/or surface number
densities of sources on the sky differ widely, more sophisticated approaches
are needed (see, e.g., Lonsdale {\it et al.} 1997, Rutledge {\it et al.} 2000,
and references therein).  One can also introduce astrophysically-motivated
constraints, e.g., flux ratios for the possible counterparts; while that may be
helpful in selecting common types of objects, it would by definition bias
against discovering sources with anomalous or unusual properties (e.g.,
spectral energy distributions).

\section{Clustering Analysis Challenges in the VO Context} 
\label{clus}  

A major exploration technique envisioned for the VO would be the unsupervised
clustering of data vectors in some parameter space of observed properties of
detected sources.  Aside from the computational challenges when large numbers
of data vectors and large numbers of dimensions are involved, this task poses
some highly non-trivial statistical and methodological problems, inviting
collaborative efforts between astronomers, computer scientists, and
statisticians.  The problems are driven not just by the sheer $size$ of the
data sets, but mainly by the {\it heterogeneity and intrinsic complexity of 
the data}. 

Separation of data into different types of objects, be it known or unknown
in nature, can be approached as a problem in automated classification or
clustering analysis.  This is a part of a more general and rapidly growing
field of Data Mining (DM) and Knowledge Discovery in Databases (KDD).  We see
here great opportunities for collaborations between astronomers and computer
scientists and statisticians.  For an overview of some of the issues and
methods, see the volume edited by Fayyad {\it et al.} (1996).

Once the measurements of source parameters from one or a number of federated
sky surveys are assembled, they can be represented as data vectors in some
(usually high-dimensionality) parameter space of observed source attributes.
The exploration of observable parameter spaces, created by combining of large
sky surveys over a range of wavelengths, will be one of the chief scientific
purposes of a VO.  This includes an exciting possibility of discovering some
previously unknown types of astronomical objects or phenomena (see Djorgovski
{\it et al.} 2001a, 2001b), and even a more general approach to SETI
(Djorgovski 2000).

Whereas some of the VO science would be done in the image (pixels) domain, and
some in the interaction between the image and catalog domains, most of the
science (at least in the initial years) may be done purely in the catalog
domain of individual or federated sky surveys.  A typical VO data set may be a
catalog of $\sim 10^8 - 10^9$ sources with $\sim 10^2$ measured attributes
each, i.e., a set of $\sim 10^9$ data vectors in a $\sim 100$-dimensional
parameter space. 

Dealing with the analysis of such data sets is obviously an inherently
multivariate statistical problem.  Complications abound: parameter correlations 
will exist; observational limits (selection effects) will generally have a
complex geometry; for some of the sources some of the measured parameters may
be only upper or lower limits; the measurement errors may vary widely; some of
the parameters will be continuous, and some discrete, or even without a
well-defined metric; etc.  In other words, analysis of the VO data sets will
present many challenging problems for applied multivariate statistics, 
and their correct statistical description should translate into a proper
(and computationally efficient) algorithmic representation.

\begin{figure}
\begin{center}
\begin{tabular}{c}
\psfig{figure=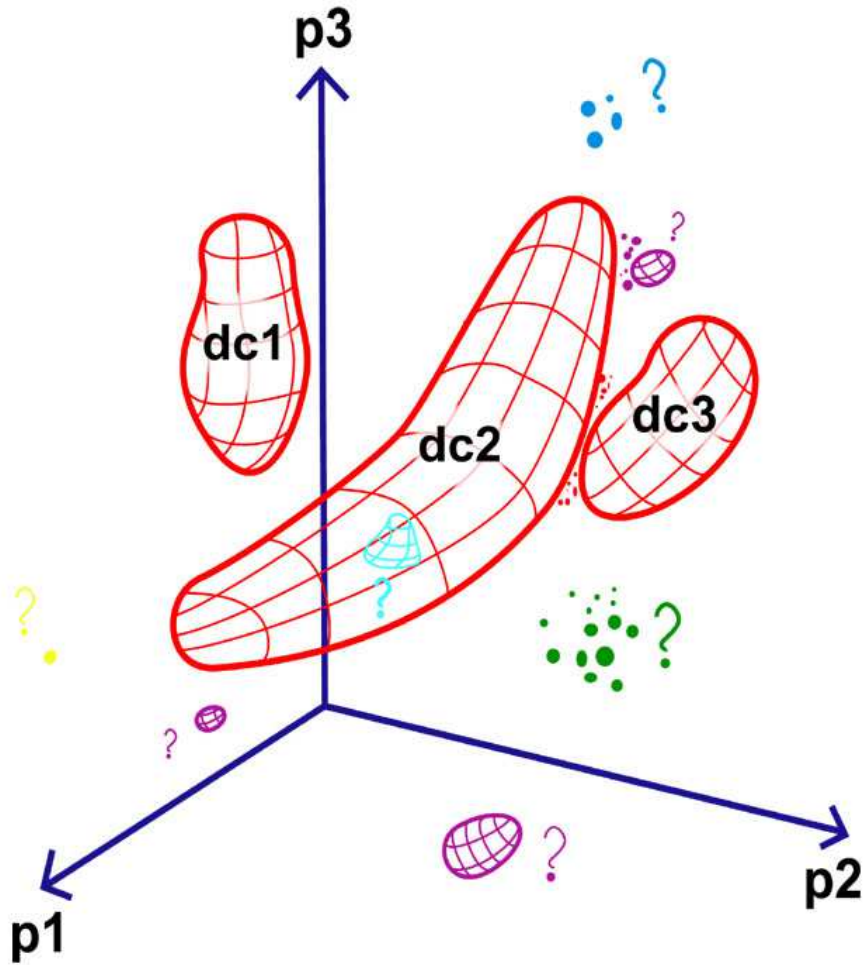,height=15cm}
\end{tabular}
\end{center}
\caption[figure 2] 
{ \label{fig:two}
A schematic illustration of the problem of clustering analysis in some
parameter space.  In this example, there are 3 dimensions, $p1$, $p2$, and $p3$
(e.g., some flux ratios or morphological paremeters), and most of the data points
belong to 3 major clusters, denoted $dc1$, $dc2$, and $dc3$ (e.g., stars,
galaxies, and ordinary quasars).  One approach is to isolate these major
classes of objects for some statistical studies, e.g., stars as probes of the
Galactic structure, or galaxies as probes of the large scale structure of the
universe, and filter out the ``anomalous'' objects.  A complementary view is to
look for other, less populated, but statistically significant, distinct
clusters of data points, or even individual outliers, as possible examples of
rare or unknown types of objects.  Another possibility is to look for holes
(negative clusters) within the major clusters, as they may point to some
interesting physical phenomenon -- or to a problem with the data.
}
\end{figure} 

\begin{figure}
\begin{center}
\begin{tabular}{c}
\psfig{figure=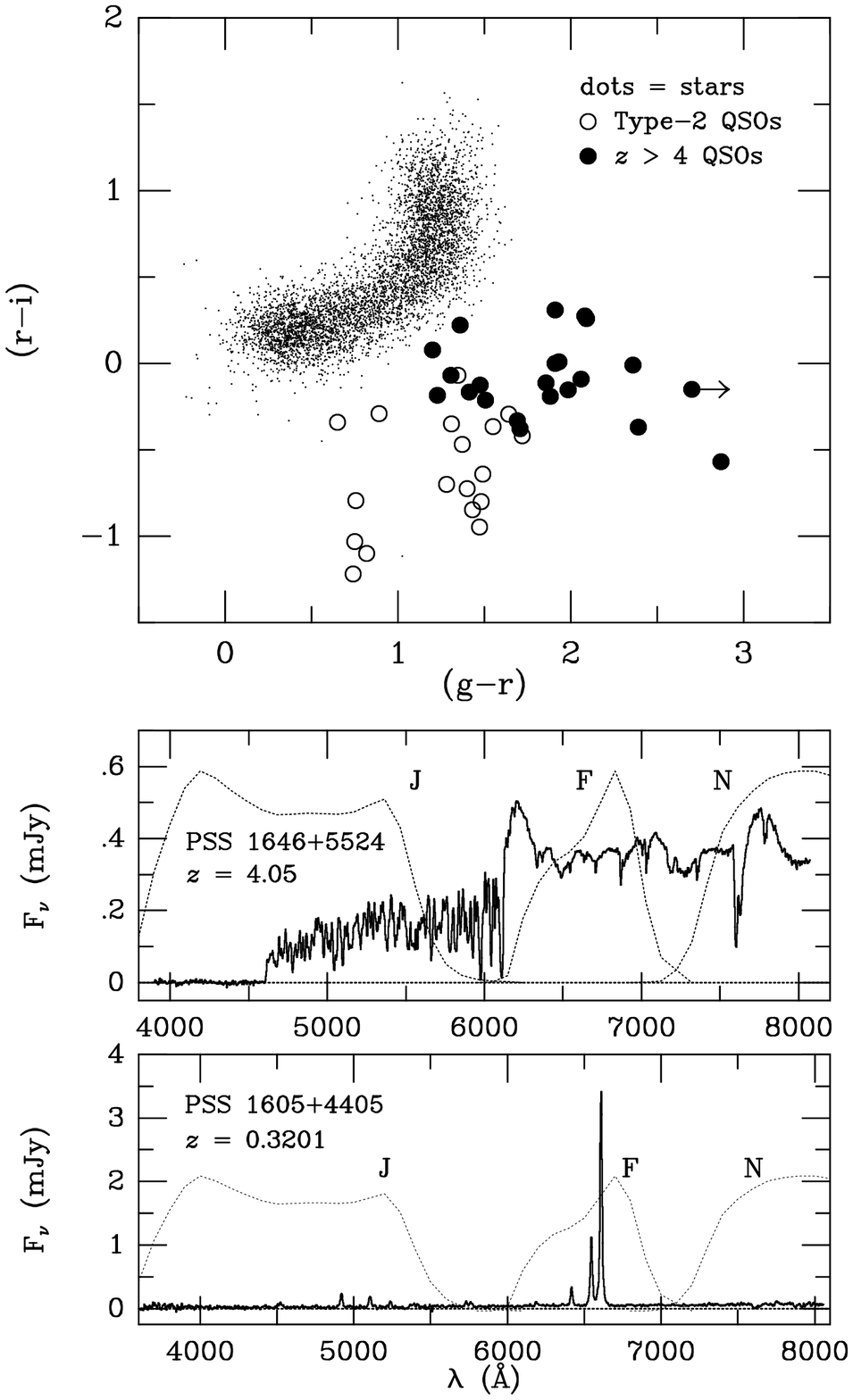,height=15cm}
\end{tabular}
\end{center}
\caption[figure 3] 
{ \label{fig:three}
An actual example of a color parameter space selection of astrophysically
interesting types of objects (high-redshift and type-2 quasars), selected from
Digital Palomar Observatory Sky Survey (DPOSS; Djorgovski {\it et al.} 1999). 
$Top:$~
A representative color-color plot for objects classified as PSF-like in DPOSS.
The dots are normal stars with $r \sim 19$ mag.  Solid circles are some of the
$z > 4$ quasars, and open circles are some of the type-2 quasars found in this
survey. 
~$Middle:$~
A spectrum of a typical $z > 4$ quasar, with the DPOSS bandpasses shown as the
dotted lines.  The mean flux drop blueward of the Ly$\alpha$ line, caused by
the absorption by Ly$\alpha$ forest and sometimes a Lyman-limit system, gives
these objects a very red $(g-r)$ color, while their intrinsic blue color is
retained in $(r-i)$, placing them in the lower right portion of this color-color
diagram.
~$Bottom:$~
A spectrum of a typical type-2 quasar, with the DPOSS bandpasses shown as the
dotted lines.  The presence of the strong [O III] lines is the $r$ band places
such objects below the stellar locus in this color-color diagram.
}
\end{figure} 

In the catalog domain, data can be viewed as a set of $n$ points or vectors in
an $m$-dimensional parameter space, where $n$ can be in the range of many
millions or even billions, and $m$ in the range of a few tens to hundreds.  The
data may be clustered in $k$ statistically distinct classes, which could be
modeled, e.g., as multivariate Gaussian clouds, and which hopefully correspond
to physically distinct classes of objects (e.g., stars, galaxies, quasars,
etc.).  If the number of object classes $k$ is known (or declared) {\it a
priori}, and training data set of representative objects is available, the
problem reduces to supervised classification, where tools such as Artificial
Neural Nets or Decision trees can be used.  This is now commonly done for
star-galaxy separation in the optical or NIR sky surveys (e.g., Odewahn 
{\it et al.} 1992, or Weir {\it et al.} 1995).  Searches for known types of
objects with predictable signatures in the parameter space (e.g., high-$z$
quasars) can be also cast in this way. 

However, a more interesting and less biased approach is where the number of
classes $k$ is not known, and it has to be derived from the data themselves.
The problem of unsupervised classification is to determine this number in some
objective and statistically sound manner, and then to associate class
membership probabilities for all objects.  Majority of objects may fall into a
small number of classes, e.g., normal stars or galaxies.  What is of special
interest are objects which belong to much less populated clusters, or even
individual outliers with low membership probabilities for any major class.
Some initial experiments with unsupervised clustering algorithms in the
astronomical context include, e.g., Goebel {\it et al.} (1989), Weir 
{\it et al.} (1995), de Carvalho {\it et al.} (1995), Yoo {\it et al.} (1996),
etc., but full-scale applications to major digital sky surveys yet remain to
be done.

This may be a computationally very expensive problem. 
For the simple $K$-means algorithm, the computing cost scales as
$K ~\times ~N ~\times ~I ~\times ~D$,
where
$K$ is the number of clusters chosen {\it a priori},
$N$ is the number of data vectors (detected objects),
$I$ is the number of iterations,
and
$D$ is the number of data dimensions (measured parameters per object).
For the more powerful Expectation Maximisation technique, the cost scales as
$K ~\times ~N ~\times ~I ~\times ~D^2$, 
and again one must decide {\it a priori} on the value of $K$.  If this number
has to be determined intrinsically from the data, e.g., with the Monte Carlo
Cross Validation method, the cost scales as
$M ~\times ~K_{max}^2 ~\times ~N ~\times ~I ~\times ~D^2$
where
$M$ is the number of Monte Carlo trials/partitions,
and
$K_{max}$ is the maximum number of clusters tried.
Even with the typical numbers for the existing large digital sky surveys
($N \sim 10^8 - 10^9$, $D \sim 10 - 100$) this is already reaching in the
realm of Terascale computing, especially in the context of an interactive and
iterative application of these analysis tools.  Development of faster and
smarter algorithms is clearly a priority.

One technique which can simplify the problem is the multi-resolution
clustering.  In this regime, expensive parameters to estimate, such as the
number of classes and the initial broad clustering are quickly estimated using
traditional techniques, and then one could proceed to refine the model locally
and globally by iterating until some objective statistical (e.g., Bayesian)
criterion is satisfied. 

Physically, the data set may consist of a number of distinct classes of
objects, such as stars (including a range of spectral types), galaxies
(including a range of Hubble types or morphologies), quasars, etc. 
Within each object class or subclass, some of the physical properties may be
correlated, and some of these correlations may be already known and some as yet
unknown, and their discovery would be an important scientific result by itself.
Some of the correlations may be spurious (e.g., driven by sample selection
effects), or simply uninteresting (e.g., objects brighter in one optical
bandpass will tend to be brighter in another optical bandpass). 
Correlations of independently measured physical parameters represent a
reduction of the statistical dimensionality in a multidimensional data
parameter space, and their discovery may be an integral part of the clustering
analysis. 

Typical scientific questions posed by an astronomer (VO user) may include:
How many statistically distinct classes of objects are in this
data set, and which objects are to be assigned to which class, along
with association probabilities?
Are there any previously unknown classes of objects, i.e.,
statistically significant ``clouds'' in the parameter space distinct
from the ``common'' types of objects (e.g., normal stars or galaxies)?
Are there rare outliers, i.e., individual objects with a low
probability of belonging to any one of the dominant classes?
Are there interesting (in general, multivariate) correlations
among the properties of objects in any given class, and what are the
optimal analytical expressions of such correlations?   

However, the rich data sets we anticipate for such studies also have many
complications.  Their construction, especially if multiple sky surveys,
catalogs, or archives are being federated (an essential VO activity) will
inevitably be imperfect, posing quality control problems which must be
discovered and solved first, before the scientific exploration even starts. 
Sources may be mismatched, there will be some gross errors or instrumental
glitches within the data, subtle systematic calibration errors may affect
pieces of the large data sets, etc. 

In addition to the size (Terabytes, billions of data vectors) and dimensionality
(hundreds of attributes per source), technical difficulties may be driven by
the intrinsic heterogeneity of the data:
Some of the parameters would be primary measurements, and others may be derived
attributes, such as the star-galaxy classification (e.g., from a supervised
classifier such as an Artificial Neural Net, or some Bayesian scheme), some may
be ``flags'' rather than numbers, some would have error-bars associated with
them, and some would not, and the error-bars may be functions of some of the
parameters, e.g., fluxes.  Some measurements would be present only as upper or
lower limits.  Some would be affected by ``glitches'' due to instrumental
problems, and if a data set consists of a merger of two or more surveys, e.g.,
cross-matched optical, infrared, and radio (and this would be a common scenario
within a VO), then some sources would be misidentified, and thus represent
erroneous combinations of subsets of data dimensions.  Surveys would be also
affected by selection effects operating explicitly on some parameters (e.g.,
coordinate ranges, flux limits, etc.), but also mapping onto some other data
dimensions through correlations of these properties; some selection effects 
may be unknown.  These issues affect the proper statistical description of the
data, which then must be reflected in the clustering algorithms. 

Even after the data sets have been optimally cleaned from glitches and fully
specified, the intrinsic nature of the object classes in the parameter space
may not be so simple.  The commonly used mixture-modeling assumption of
clusters represented as multivariate Gaussian clouds is rarely a good
descriptor of the reality.  These multivariate ``clouds'' in the parameter
space may have a power-law or exponential tails in some or all of the
dimensions, and some may have sharp cutoffs, etc. 
This becomes a critical issue in evaluating the membership probabilities in
partly overlapping clusters, or in a search for outliers (anomalous events) in
the tails of the distributions.  In general, the proper functional forms for
the modeling of clusters are not known {\it a priori}, and must be discovered
from the data.  Applications of non-parametric techniques may be essential
here. 

\begin{figure}
\begin{center}
\begin{tabular}{c}
\psfig{figure=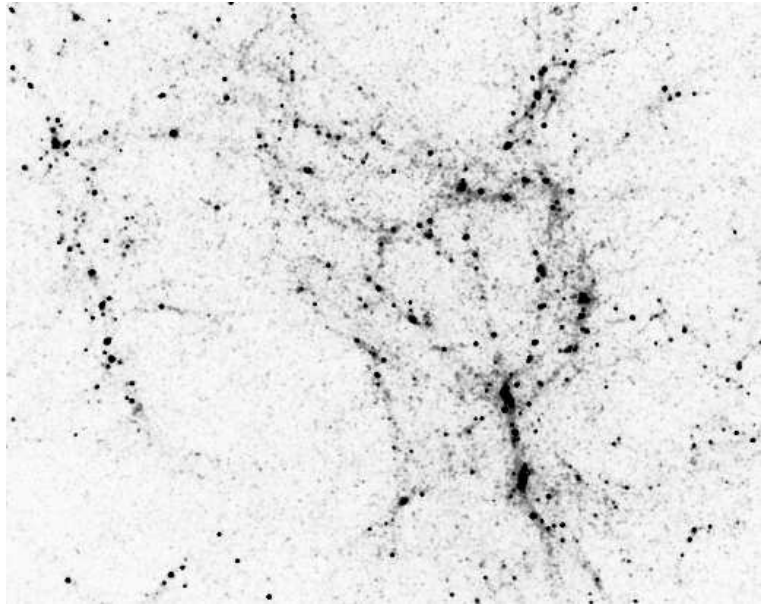,height=8cm}
\end{tabular}
\end{center}
\caption[figure 4] 
{ \label{fig:four}
Clustering of galaxies in the physical space (a special case of data vectors
in a parameter space, whose axes are simply spatial coordinates) illustrates 
some of the problems described here.  Actual galaxy clusters are non-linear
overdensities superposed on a clustered background; their detection requires
an algorithm which can separate them correctly from fluctuations in an
$\sim 1/f$ noise background.  The clustering has a complex topology, of
filaments, sheets, and voids (negative clusters), and most commonly used
algorithms would simply cut up this density field into strings of Gaussian
beads, thus missing the essential feature of the physical distribution.
(Image from a numerical simulation of structure formation, by F.~Summers and
the GC3 consortium, http://lca.ncsa.uiuc.edu:8080/Summers/).
}
\end{figure} 

The clusters may be well separated in some of the dimensions, but not in others.
How can we objectively decide which dimensions are irrelevant, and which ones
are useful?  An automated and objective rejection of the ``useless'' dimensions,
perhaps through some statistically defined enthropy criterion, could greatly
simplify and speed up the clustering analysis. 
Techniques for dimensionality reduction, including principal component analysis
and others can be used as preprocessing techniques to automatically derive the
dimensions that contain most of the relevant information. 

The {\it topology} of clustering may not be simple:  there may be clusters
within clusters, holes in the data distribution (negative clusters?),
multiply-connected clusters, etc. 
Clustering may be hierarchical or multi-scale (i.e., clusters embedded within
the clusters), etc. 

In many situations, scientifically informed input is needed in designing the
clustering experiments.  Some observed parameters may have a highly
significant, large dynamical range, dominate the sample variance, and naturally
invite division into clusters along the corresponding parameter axes; yet they
may be completely irrelevant or uninteresting scientifically.  For example, if
one wishes to classify sources of the basic of their broad-band spectral energy
distributions (or to search for objects with unusual spectra), the mean flux
itself is not important, as it mainly reflects the distance; coordinates on the
sky may be unimportant (unless one specifically looks for a spatial
clustering); etc.  Thus, a clustering algorithm may divide the data set
along one or more of such axes, and completely miss the really scientifically
interesting partitions, e.g., according to the colors of objects.

One can also use intelligent sampling methods where one forms ``prototypes'' of
the case vectors and thus reduces the number of cases to process.  Prototypes
can be determined from simple algorithms to get a rough estimate, and then
refined using more sophisticated techniques.  A clustering algorithm can
initially operate in the prototype data space.  The clusters found can then be
refined by locally replacing each prototype by its constituent population and
reanalyzing the clusters.

\section{Concluding Comments} 
\label{concl}  

An array of good unsupervised classification techniques will be an essential
part of a VO toolkit.  Blind applications of the commonly used (commercial or
home-brewed) clustering algorithms could produce some seriously misleading or
simply wrong results.  The clustering methodology must be robust enough to cope
with these problems, and the outcome of the analysis must have a solid
statistical foundation. 

Effective and powerful data visualization, applied in the parameter space
itself, is another essential part of the interactive clustering analysis.  
Good visualisation tools are also critical for the interpretation of results,
especially in an iterative environment.  While clustering algorithms can assist
in the partitioning of the data space, and can draw the attention to anomalous
objects, ultimately a scientist guides the experiment and draws the conclusions.

Another key issue is interoperability and reusability of algorithms and models
in a wide variety of problems posed by a rich data environment such as
federated digital sky surveys in a VO.  Implementation of clustering analysis
algorithms must be done with this in mind.

Finally, scientific verification and evaluation, testing, and follow-up on
any of the newly discovered classes of objects, physical clusters discovered by
these methods, and other astrophysical analysis of the results is essential in
order to demonstrate the actual usefulness of these techniques for a VO or
other applications.  Clustering analysis can be seen as a prelude to the more
traditional type of astronomical studies, as a way of selecting of interesting
objects of samples, and hopefully it can lead to advances in statistics and
applied computer science as well.

In our experience, design and application of clustering algorithms must involve
close working collaboration between astronomers, computer scientists and
statisticians.  There are too many unspoken assumptions, historical background
knowledge specific to a given discipline, and an opaque jargon; constant
communication and interchange of ideas are essential.  Such genuine intellectual
partnerships are guaranteed to produce results and advances in all of the
disciplines involved. 


\acknowledgments     
This work was supported in part by the NASA AISRP grant NAG5-9482, and by
private foundations.  We wish to thank numerous collaborators, and colleagues
working on making the VO vision become a reality. 


\def\pp{\noindent\parshape 2 0.0 truecm 17.5 truecm 1.0 truecm 16.5 truecm}

\section{References}       

\pp Banday, A.J. {\it et al.} (eds.) 2001, {\sl Mining the Sky}, 
    ESO Astrophysics Symposia, Berlin: Springer Verlag, in press

\pp Brunner, R., Djorgovski, S.G., \& Szalay, A.S. (eds.) 2001a,
    {\sl Virtual Observatories of the Future}, 
    {\it A.S.P.~Conf.~Ser.} vol. {\bf 225}
    [http://www.astro.caltech.edu/nvoconf/]

\pp Brunner, R.J., Djorgovski, S.G., Prince, T.A., \& Szalay, A. 2001b,
    in: {\sl Handbook of Massive Data Sets}, eds. J. Abello, P. Pardalos, 
    \& M. Resende, in press
    [http://www.arXiv.org/abs/astro-ph/0106481]

\pp de Carvalho, R., Djorgovski, S., Weir, N., Fayyad, U., Cherkauer, K.,
    Roden, J., \& Gray, A. 1995, 
    in: {\sl Astronomical Data Analysis Software and Systems IV}, 
    eds. R. Shaw, H. Payne, \& J. Hayes,
    {\it A.S.P.~Conf.~Ser.} {\bf 77}, 272

\pp Djorgovski, S.G., Gal, R.R., Odewahn, S.C., de Carvalho, R.R., 
    Brunner, R.J., Longo, R., \& Scaramella, R. 1998, 
    in: {\sl Wide Field Surveys in Cosmology}, eds. S. Colombi {\it et al.}, 
    Gif sur Yvette: Eds. Fronti\`eres, p. 89
    [http://www.arXiv.org/abs/astro-ph/9809187]

\pp Djorgovski, S.G., Mahabal, A., Brunner, R., Gal, R.R., Castro, S., 
    de Carvalho, R.R., \& Odewahn, S.C. 2001, 
    in: {\sl Virtual Observatories of the Future}, 
    eds. R.~Brunner, S.G.~Djorgovski \& A.~Szalay, 
    {\it A.S.P.~Conf.~Ser.} {\bf 225}, 52
    [http://www.arXiv.org/abs/astro-ph/0012453]

\pp Djorgovski, S.G., Brunner, R.J., Mahabal, A.A., Odewahn, S.C., de Carvalho,
    R.R., Gal, R.R., Stolorz, P., Granat, R., Curkendall, D., Jacob, J., \& 
    Castro, S. 2001,
    in: {\sl Mining the Sky}, eds. A.J.~Banday {\it et al.}, ESO Astrophysics 
    Symposia, Berlin: Springer Verlag, in press
    [http://www.arXiv.org/abs/astro-ph/0012489]

\pp Djorgovski, S.G. 2000, 
    in: {\sl Bioastronomy '99}, eds. G. Lemarchand \& K. Meech, 
    {\it A.S.P.~Conf.~Ser.} {\bf 213}, 519

\pp Fayyad, U., Piatetsky-Shapiro, G., Smyth, P., \& Uthurusamy, R. (eds.) 1996,
    {\sl Advances in Knowledge Discovery and Data Mining}, 
    Boston: AAAI/MIT Press

\pp Goebel, J., Volk, K., Walker, H., Gerbault, F., Cheeseman, P., Self, M.,
    Stutz, J., \& Taylor, W. 1989, A\&A, 222, L5 

\pp Harwit, M. 1975, QJRAS, 16, 378

\pp Harwit, M., \& Hildebrand, R. 1986, Nature, 320, 724

\pp Lonsdale, C., {\it et al.} 1997,
    in: {\sl New Horizons From Multi-Wavelength Sky Surveys}, IAU Symp. 179,
    eds. B. McLean {\it et al.}, p. 450, Dordrecht: Kluwer

\pp McKee, C., Taylor, J., {\it et al.} 2001,
    {\sl Astronomy and Astrophysics in the New Millennium},
    Washington, DC: National Academy Press
    [http://www.nap.edu/books/0309070317/html/]

\pp Odewahn, S.C., Stockwell, E., Pennington, R., Humphreys, R., \& Zumach, W.
    1992, AJ, 103, 318 

\pp Paczy\'nski, B. 2000, PASP, 112, 1281

\pp Rutledge, R., Brunner, R., Prince, T., \& Lonsdale, C. 2000, ApJSS, 131, 335

\pp Weir, N., Fayyad, U., \& Djorgovski, S. 1995, AJ, 109, 2401

\pp White Paper 2001, ``Toward a National Virtual Observatory: 
    Science Goals, Technical Challenges, and Implementation Plan'', 
    in: {\sl Virtual Observatories of the Future}, 
    eds. R. Brunner, S.G. Djorgovski, \& A. S. Szalay, 
    {\it A.S.P.~Conf.~Ser.} {\bf 225}, p.~353.
    [http://www.arXiv.org/abs/astro-ph/0108115]

\pp Yoo, J., Gray, A., Roden, J., Fayyad, U., de Carvalho, R., \& 
    Djorgovski, S. 1996, 
    in: {\sl Astronomical Data Analysis Software and Systems V}, 
    eds. G. Jacoby \& J. Barnes, 
    {\it A.S.P.~Conf.~Ser.} {\bf 101}, 41

\pp Zwicky, F. 1957, {\sl Morphological Astronomy}, Berlin: Springer Verlag 

\end{document}